\documentstyle[psfig]{l-aa}

\newcommand{\iK}{\mbox{{\scriptsize K}}}
\newcommand{\iV}{\mbox{{\scriptsize V}}}

\begin{document}

   \thesaurus{06         
                                          evolution of stars
	     (08.04.1;   
              08.06.3;   
	      08.22.3;   
              10.19.3)}  
   \title{Oxygen-rich Mira variables: near-infrared luminosity 
          calibrations
   \thanks{Based on data from the HIPPARCOS astrometry satellite}}

   \subtitle{Populations and period--luminosity relations}

   \author{R.\ Alvarez
	   \inst{1},
           M-O.\ Mennessier
	   \inst{1},
           D.\ Barth\`es
	   \inst{1},
           X.\ Luri
	   \inst{2,}\inst{3},
           and J.A.\ Mattei
           \inst{4}
	  }

   \offprints{alvarez@graal.univ-montp2.fr}

   \institute{GRAAL,
	      Universit\'e Montpellier II,
              UPRESA 5024/CNRS,
	      F--34095 Montpellier Cedex 05, France
	      \and
	      Departament d'Astronomia i Meteorologia,
              Universitat de Barcelona,
              Avda. Diagonal 647,
              E--08028 Barcelona, Spain
	      \and
              DASGAL,
              Observatoire de Paris-Meudon,
              F--92195 Meudon Cedex, France
              \and
              AAVSO,
              25 Birch Street,
              Cambridge,
              Massachusetts 02138--1205, USA
	      }

   \date{Received 30 April 1997/ Accepted 11 July 1997}

   \maketitle

   \markboth{Alvarez et al.: Oxygen-rich Mira variables: near-infrared 
             luminosity calibrations}
            {Period--luminosity relations}

   \begin{abstract}
    HIPPARCOS astrometric and kinematical data of oxygen-rich Mira 
    variables are used to calibrate absolute near-infrared magnitudes 
    and kinematic parameters.
    Two sets of near-infrared magnitudes compiled from different 
    authors are used: broad-band K and narrow-band photometric 
    measurements at 1.04 $\mu$m (104 filter).
    Three distinct classes of stars with different kinematics and
    scale height have been identified. The two most significant 
    groups present characteristics close to the ones usually assigned
    to extended/thick disk--halo population and old disk population 
    respectively, and thus they might differ by their metallicity 
    abundance.
    They exhibit different period distributions, as expected if these
    two groups actually correspond to populations of distinct initial 
    masses, ages and metallicities.
    Two parallel period--luminosity relations are found in K as well
    as in 104, one for each significant population. The shift between 
    these relations is interpreted as the consequence of the effects 
    of metallicity abundance on the luminosity.
    
   \end{abstract}


\section{Introduction}

Mira variables, due to their intrinsic brightness and large range
of their ages, mark a unique stage in stellar evolution of 
intermediate-mass stars and thus are important in the study of 
stellar populations in our Galaxy.
Knowledge of their distances is crucial to understand the Galactic 
structure evolution as well as the pulsational properties of these 
stars.
The existence of infrared and bolometric period--luminosity (PL)
relations in the Large Magellanic Cloud (LMC) for Mira variables
(see, e.g., Feast et al.\ 1989) has allowed us to estimate Galactic
distances for a large number of Miras (Jura \& Kleimann 1992; 
Jura et al.\ 1993; Alvarez \& Mennessier 1997). 
But such works have been limited by the (unavoidable untill recently) 
assumption on the choice of the zero point for the Galactic Mira 
PL relation. Now, the release of HIPPARCOS data enables one to 
proficiently investigate this particular point.\\
The results presented in this paper constitute the application of the 
LM (Luri, Mennessier et al.\ 1996a; hereafter Paper I) method to 
HIPPARCOS data concerning oxygen-rich Miras. 
This method is based on a maximum-likelihood estimation using 
apparent magnitudes, trigonometrical parallaxes, proper motions and 
radial velocities. It has been applied to two different samples of
about one hundred oxygen-rich Miras for which two sets of 
near-infrared (K and 104) magnitudes at maximum have been compiled 
from different authors.
These apparent magnitudes complement the kinematical data: 
the trigonometric parallaxes and proper motions are obtained from the 
recently available HIPPARCOS Catalogue (ESA 1997) and the radial 
velocities from the HIPPARCOS Input Catalogue (Turon et al.\ 1992).\\
A preliminary approach to the fundamental problem of absolute 
magnitudes and distances determinations was made by Luri et al.\ 
(1996b) using HIPPARCOS Input Catalogue kinematics data and visual 
magnitudes.
Infrared data are maybe more suitable to study the very cool stars
as they emit most of their energy in the infrared.
At least in K band, these measurements are less sensitive than visual 
magnitudes to interstellar and circumstellar extinction. The effect of 
molecular absorption, which considerably varies in visual region over 
a cycle, is also weaker in K or 104. Furthermore, the 
existence of period--luminosity relations is well attested in infrared 
bands. All these properties contribute to the interest of applying the 
LM method to near-infrared data.

\section{The samples and data}

In this paper, we focus on oxygen-rich Miras with available data in 
HIPPARCOS Catalogue: they constitute a first sample of nearly 200 
objects. Only part of them have been observed in near-infrared.
As a maximum-likelihood method gives statistical results, the need 
for a sufficiently large working sample is therefore critical. Thus, 
we decided to also include oxygen-rich semi-regular (SR) variables in
order to produce more reliable estimations. The SR variables are very 
similar to the Miras: they are arbitrarily discriminated from Miras 
according to their smaller visual amplitude, but, as observed in the
LMC (Hughes et al.\ 1990), their absolute magnitude distributions
should be similar. We have checked a posteriori that the oxygen-rich SR 
and the M--Miras of our sample actually exhibit similar kinematics and
luminosities. The so-defined sample is thus kinematically homogeneous 
so that the LM method could be used with benefit. SR variables will be 
studied apart in a forthcoming paper.
The trigonometric parallaxes and proper motions of the initial sample
are obtained from the HIPPARCOS Catalogue (ESA 1997). They are 
complemented with radial velocities (HIPPARCOS Input Catalogue, Turon
et al.\ 1992) and with:
\begin{enumerate}
  \item {\bf apparent K magnitudes} at maximum luminosity compiled 
  from different sets of K--band observations available in the 
  literature: 
  Catchpole et al.\ (1979) presented a total set of 2883 JHKL 
  measurements of 223 Mira-type variables and 404 late-type stars of 
  various types; 
  Fouqu\'e et al.\ (1992) published JHKLM photometry of 516 
  sources pertaining to the IRAS Point Source Catalogue; 
  Whitelock et al.\ (1994) gave over 1500 JHKL photometry observations 
  for 61 Miras; 
  Kerschbaum \& Hron (1994) and Kerschbaum (1995) presented JHKL'M 
  observations of respectively 200 and 44 SR variables.
  \item {\bf apparent 104 magnitudes} at maximum luminosity. The 104 
  filter forms part of a five-colour narrow-band photometric system 
  used by Lockwood (1972). He observed 281 M-- and S--type Mira 
  variables and 11 SR stars for a total of nearly 1800 sets of 
  measurements. 
  The five-colour photometric system is based on Wing's 27 colours 
  system (Wing 1967). The peak wavelength of the 104 filter is 
  10351 \AA, and the half-power bandwidth is 125 \AA. This narrow-band
  filter matches a region relatively free of molecular absorption 
  ( Wing 1967; Alvarez \& Plez 1997) and hence can be considered as a 
  reliable measurement of "continuum". 
  Furthermore, the distances that can be derived from apparent 104 
  magnitudes are in good agreement with those obtained using apparent K 
  magnitudes (Alvarez \& Mennessier 1997). 
  The 104 filter can be regarded as meaningful as K band in the study 
  of these late-type stars.
  From Lockwood's sample, only stars with at least one observation
  in the phase range 0.8--0.2 (i.e.\ near maximum luminosity) have been 
  kept before determining the brightest 104 value.
\end{enumerate}

The two final samples are coincidentally both composed of 103 Mira 
variables, plus 129 SR and 8 SR for the K and 104 sample respectively.
Seventy variables (64 Miras and 6 SR) belong to both of them.

\section{The method}

The reader is referred to Paper I for a thorough description of the 
LM method. We outline here its most important features. 
This method, based on the maximum-likelihood principle, allows us 
to simultaneously calibrate the luminosity and determine the mean 
kinematic characteristics and spatial distribution of a given sample.
This sample is specifically modeled with appropriate distribution
functions corresponding to the absolutes magnitudes, kinematics and
spatial distribution. 
Sampling effects, the galactic differential rotation and observational 
errors are rigorously taken into account by including appropriate
functions in the density law describing the sample.
The effects of the observational errors in apparent magnitude are
neglected, and only the errors in trigonometrical parallax, proper 
motions and radial velocity are included in the density law.
The method is able to use inhomogeneous samples, i.e.\ samples 
composed of a mixture of groups of stars with different luminosity, 
kinematics or spatial distribution. In this case the method identifies 
and separates the groups. Moreover, the LM method assigns each star
to a group and estimates its most probable distance.\\
In view to model each group of stars, the following distribution 
functions have been adopted:
\begin{enumerate}
  \item {\bf Distribution of absolute infrared magnitudes:} a gaussian
  law with mean $M_0$ and standard deviation $\sigma_M$
  \item {\bf Velocity distribution:} a Schwarzschild ellipsoid with means
  ($U_0,V_0,W_0$) and dispersions ($\sigma_U,\sigma_V,\sigma_W$)
  \item {\bf Spatial distribution:} an exponential disc with scale height 
  $Z_0$
\end{enumerate}
These parameters are determined by the LM method at the same time as 
the relative proportion of each group.   

\section{Populations separation}

\subsection{Absorption correction in 104}

The interstellar absorption is taken into account by the LM
method in the determination of the parameters.
Fluks et al.\ (1994) have tabulated an extended mean extinction law 
based on the observed mean extinction law of Savage \& Mathis (1979) 
and on the theoretical extinction law of Steenman \& Th\'e (1989, 1991).
From the tabulated $a(\lambda)/E_{B-V}$ values, it appears that the 
extinction in 104 is smaller than in V by a factor 0.37, which is
non-negligible.\\ 
The absorption correction in 104 is done by using a detailed model
of visual interstellar absorption, based on the scheme of Arenou et al.\
(1992), scaled by the factor 0.37.
A null absorption correction is assumed for the K magnitudes. 
As quoted by Allen (1973), absorption is less pronounced in K than in 
V by at least a factor ten and thus it should be negligible. 
The validity of this particular point will be checked later.

\subsection{Number of groups and mean parameters}

Wilks test indicates that the three group solution is the optimal one
for both samples.
Tables~1 and 2 give the maximum-likelihood estimates of the parameters
for the K and the 104 sample respectively.
In these tables, the estimates of the parameters are given in the
columns marked $\theta$  and the corresponding errors are given in
the columns marked $\sigma$. These errors were calculated using
Monte-Carlo simulations: simulated samples were generated and
LM estimations were performed with them. The dispersion of these
estimates was taken as the error on the results.\\
The two sets of kinematic parameters are in good agreement given
the estimation errors. 
It is extremely satisfactory to obtain such a good agreement despite 
the uncertainties inherent to any statistical method, and despite the 
fact that the majority of both samples differ.
Among the 70 common variables, only 8 are classified in discrepant 
groups.

\begin{table}
\caption[]{Model parameters using $m_{\iK}$ 
           (232 variables including 103 Miras)}
\begin{flushleft}
\begin{tabular}{llllllllll}
\noalign{\smallskip}
\hline
\noalign{\smallskip}
  &  & \multicolumn{2}{l}{group 1} &  & \multicolumn{2}{l}{group 2} 
  &  & \multicolumn{2}{l}{group 3} \\
\cline{3-4} \cline{6-7} \cline{9-10}
  &  & $\theta$ & $\sigma$ &  & $\theta$ & $\sigma$ &  & $\theta$ & $\sigma$ \\ 
\noalign{\smallskip}
\hline
\noalign{\smallskip}
$M_0$      & (mag)          & -6.3 & 0.7 &  & -6.1 & 1.6 &  & -6.7 & 0.9 \\
$\sigma_M$ & (mag)          &  1.0 & 0.4 &  &  0.4 & 0.4 &  &  0.8 & 0.5 \\
$U_0$      & (km.s$^{-1}$)  &  -11 &   6 &  &  -53 &  17 &  &  -33 &  40 \\
$\sigma_U$ & (km.s$^{-1}$)  &   37 &   8 &  &    1 &   7 &  &   93 &  34 \\
$V_0$      & (km.s$^{-1}$)  &  -23 &   6 &  &  -57 &  74 &  &  -93 &  53 \\
$\sigma_V$ & (km.s$^{-1}$)  &   22 &   4 &  &   15 &  30 &  &   75 &  24 \\
$W_0$      & (km.s$^{-1}$)  &  -12 &   4 &  &  -33 &  10 &  &   -2 &  33 \\
$\sigma_W$ & (km.s$^{-1}$)  &   20 &   5 &  &    3 &   4 &  &   58 &  31 \\
$Z_0$      & (pc)           &  260 &  40 &  &  370 & 180 &  &  820 & 240 \\
\%         &                &   81 &   7 &  &    2 &   1 &  &   17 &   7 \\
\noalign{\smallskip}
\hline
\end{tabular}
\end{flushleft}
\end{table}

\begin{table}
\caption[]{Model parameters using $m_{104}$ 
           (111 variables including 103 Miras).}
\begin{flushleft}
\begin{tabular}{llllllllll}
\noalign{\smallskip}
\hline
\noalign{\smallskip}
  &  & \multicolumn{2}{l}{group 1} &  & \multicolumn{2}{l}{group 2} 
  &  & \multicolumn{2}{l}{group 3} \\
\cline{3-4} \cline{6-7} \cline{9-10}
  &  & $\theta$ & $\sigma$ &  & $\theta$ & $\sigma$ &  & $\theta$ & $\sigma$ \\ 
\noalign{\smallskip}
\hline
\noalign{\smallskip}
$M_0$      & (mag)          & -5.3 & 0.4 &  & -5.5 & 1.5 &  & -5.6 & 0.9 \\
$\sigma_M$ & (mag)          &  0.9 & 0.2 &  &  1.0 & 0.9 &  &  1.0 & 0.5 \\
$U_0$      & (km.s$^{-1}$)  &  -14 &   8 &  &  -51 &  12 &  &  -32 &  40 \\
$\sigma_U$ & (km.s$^{-1}$)  &   45 &  13 &  &    2 &   2 &  &  117 &  45 \\
$V_0$      & (km.s$^{-1}$)  &  -21 &   7 &  &  -41 &   9 &  & -123 &  42 \\
$\sigma_V$ & (km.s$^{-1}$)  &   25 &   7 &  &    5 &   2 &  &   41 &  31 \\
$W_0$      & (km.s$^{-1}$)  &   -9 &   4 &  &  -34 &  19 &  &   -8 &  37 \\
$\sigma_W$ & (km.s$^{-1}$)  &   20 &   9 &  &   14 &   4 &  &   75 &  37 \\
$Z_0$      & (pc)           &  400 & 130 &  & 160  &  80 &  & 2630 & 1020 \\
\%         &                &   73 &   8 &  &    6 &   3 &  &   21 &   8 \\
\noalign{\smallskip}
\hline
\end{tabular}
\end{flushleft}
\end{table}

Two significant populations are well separated. 
Group~1, which is the main one (about 75 \% of the sample), 
has the kinematics of late disk stars. The scale height is 
characteristic of the old disk population (it is perhaps slightly 
too large for the 104 sample). This group can be interpreted as 
the standard population which has a exponential scale height of 
$\sim 300$ pc (Jura \& Kleinmann 1992).
Group~3 (about 20 \% of the samples) has a larger velocity 
ellipsoid. The scale height is much more important. The very large 
scale height in Table 2 is an artifact of the method: it only means a 
spherical spatial distribution. 
The large velocity ellipsoid and the important scale height 
characterizes a population older than the group 1. 
They might belong to the extended/thick (E/T) disk or they might be 
halo stars.\\
The mean period of the Mira variables belonging to group~1 is 321~d 
for the K sample and 308~d for the 104 sample. It is respectively 
217~d and 216~d for the Miras of group~3. Both distributions are
overlapping.
The existence of different period distributions was already attested 
by Jura \& Kleinmann (1992) and certainly reflects the different 
evolutionary paths followed by two populations of distinct initial 
masses, ages and metallicities.\\
Group~2 is a very small group with very low velocity dispersion. 
They might form a sub-population of younger stars. The small 
number of stars prevents further interpretation.\\

The present results are in good agreement with those obtained 
by Luri et al.\ (1996b) for a sample of 90 M--Mira variables using 
apparent {\it visual} magnitudes and HIPPARCOS Input Catalogue data 
(Turon et al.\ 1992). In particular, the three present groups were
already attested. This confirms the confidence that we can have in
the results, since the data differ by the samples size, the 
apparent magnitudes, the source and the accuracy of the proper 
motions and the introduction of the parallaxes.\\
Comparaison has also been performed with the results of 
Mennessier et al.\ (1997a, hereafter Paper II; 1997c) who have 
applied the LM method to the total sample of Long Period Variables 
(LPVs), i.e.\ Mira, SR and Irregular variables, of M, S and C type 
(nearly 900 stars), using apparent visual magnitudes and 
HIPPARCOS data (ESA 1997).
As this total sample is very large and its observational selection
criteria (Mennessier \& Baglin 1988) are well defined, the kinematic
parameters and the population separation obtained in Paper II should 
be considered as the most reliable among all the applications of the 
LM method to sub-samples of LPVs.
The results of Paper II show a separation into several groups.
Indeed, due to the larger number of stars, the partition into the 
different groups is more accurate than ours and several populations 
can be defined: they represent the gradual transition between the 
younger disk population and the stars clearly belonging to the halo. 
A specific look at the oxygen-rich Miras confirms what we find in 
the present work: they mainly (79 \%) belong to the groups that 
correspond to the old disk population, while a non-negligible 
fraction (12\%) are found in the groups that might be assimilated 
to the E/T disk and the halo. Some M--Miras (9 \%) belong to a 
younger disk population.

\section{Comparison of distance estimates}  

Once a star is assigned to a group, the LM method enables us, via the
distance marginal density law, to estimate the most probable value of 
the distance and its error. For each star a distance 
$r_{\iK}$ and/or a distance $r_{104}$ is therefore proposed.
The distance estimates using the apparent V magnitudes (Paper II), 
$r_{\iV}$, are compared to the distances derived from the K and 104 
magnitudes in Fig(s).~1 and 2 respectively.
The stars are denoted by symbols indicating the population to which
they belong {\it as assigned by the classification of Paper II}, i.e.\ 
younger disk population (asterisks), old disk population (open circles)
or E/T disk and the halo population (filled circles).

\begin{figure}
  \centerline{\psfig{figure=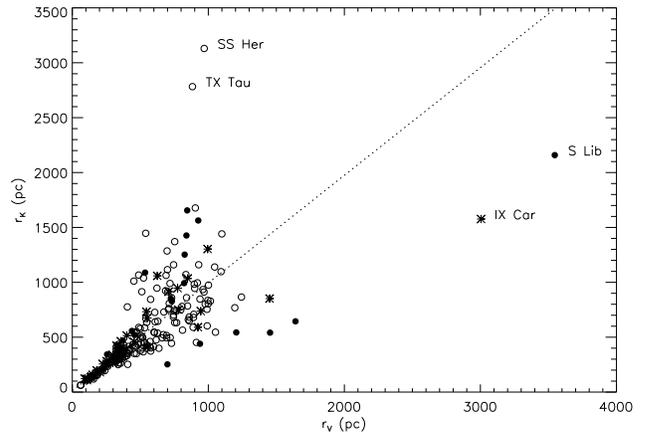,width=3.4in}}
  \caption{Comparison of the distance estimates using the apparent K
   magnitudes (this work) and those using the apparent V magnitudes
   (Mennessier et al.\ 1997a,c). Asterisks represent LPVs belonging to 
   the younger disk population, open circles to the old disk population 
   and filled circles to the E/T disk or halo population, as determined 
   by the classification of Mennessier et al.\ (1997a,c). The dotted line 
   is the regression line}
\end{figure}

\begin{figure}
  \centerline{\psfig{figure=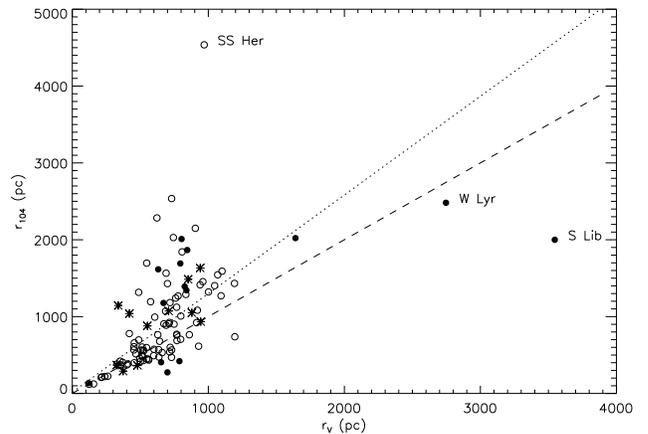,width=3.4in}}
  \caption{Comparison of the distance estimates using the apparent 104
   magnitudes (this work) and those using the apparent V magnitudes
   (Mennessier et al.\ 1997a,c). Same symbols as Fig.~1. Dotted and dashed
   lines are the regression and one-to-one lines respectively}
\end{figure}

The distances $r_{\iK}$ are in good agreement with the distances
$r_{\iV}$, considering the unavoidable uncertainties which result 
from the statistical character of the estimating method and increase 
with distance.
A least-squares linear fit gives $r_{\iK} \approx 0.98 r_{\iV}$.
The regression line is indicated in Fig.~1.
It is worth remarking that no systematic discrepancy appears between 
the two sets of distances.
This confirms that the extinction in K is very small.
The outlying stars are indicated with their name. 
Assignation to different populations in V and in K explains these 
discrepancies. Indeed, the group assignation is only a statistical 
result and some inconsistencies remain.\\
For the 104 distances, a least-squares linear fit gives 
$r_{104} \approx 1.28 r_{\iV}$.
As in K and for the same reason, there are some outlying stars.
The systematic discrepancy might be due to sampling effects 
generating a bias. Indeed, due to the small number of objects, the 
sample of 104 observations is probably not well representative of 
the HIPPARCOS sampling function adopted in our calculations. As a 
consequence, the 104 distances and magnitudes are probably less 
reliable than the K ones. 
Nevertheless, it is remarkable that the noticeable results from K 
presented in Sect.~4 (separation into populations of different 
kinematics and period distribution) and those of Sect.~6 (existence 
of two nearly parallel period--luminosity relations) still hold 
from 104.

\section{Mira period--luminosity relations}

\subsection{Period--luminosity relations and populations}

The individual absolute magnitudes of Miras can also be obtained with 
the LM method. They are plotted against the logarithm of the periods 
in Fig(s).~3 and 4, which correspond to the K and 104 samples 
respectively. Periods were taken from the General Catalogue of 
Variable Stars (Kholopov 1985, 1987) or from 
Mennessier et al.\ (1997b). The three groups that we defined in the 
present work are distinguished by different symbols. For the two most 
significant groups (1 and 3), least-square linear fits are obtained.
The different equations of the regression lines for our HIPPARCOS 
samples are:
\begin{enumerate}
  \item In K band:
  \begin{enumerate}
    \item Group 1 (85 Miras)
      \begin{equation}
        M_{\iK} = 0.976 - 3.41 \log P , \\
      \sigma = 0.72
      \end{equation}
    \item Group 3 (16 Miras)
      \begin{equation}
        M_{\iK} = -0.129 - 3.18 \log P , \\
      \sigma = 0.52
      \end{equation}
  \end{enumerate}
  \item In 104 filter:
  \begin{enumerate}
    \item Group 1 (76 Miras)
      \begin{equation}
        M_{104} = -0.755 - 2.19 \log P , \\
      \sigma = 0.54
      \end{equation}
    \item Group 3 (19 Miras)
      \begin{equation}
        M_{104} = -0.804 - 2.52 \log P , \\
      \sigma = 0.67
      \end{equation}
  \end{enumerate}
\end{enumerate}

\begin{figure}
  \centerline{\psfig{figure=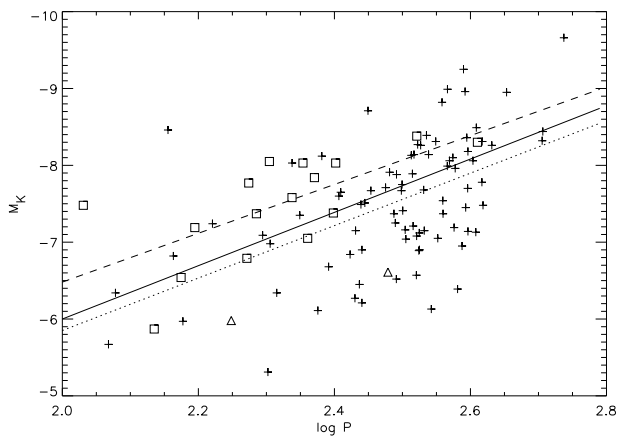,width=3.4in}}
  \caption{Period--luminosity relations in K band. Crosses represent
   Miras belonging to group~1, triangles to group~2 and squares to 
   group~3. Dotted line and dashed line are the PL fit relations 
   for group~1 and 3 respectively. Solid line is the PL relation 
   determined by van Leeuwen et al.\ (1997)}
\end{figure}

\begin{figure}
  \centerline{\psfig{figure=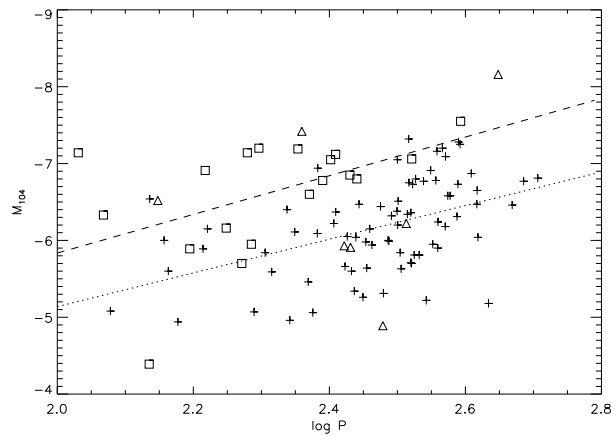,width=3.4in}}
  \caption{Period--luminosity relations in 104 filter. Same symbols
   as Fig.~3}
\end{figure}

The period--luminosity relations determined for groups~1 and 3, 
in K as well as in 104, are remarkably parallel, despite the large 
scatter about the regression lines. The different period 
distributions of the groups do not prevent from obtaining similar 
slopes. It is worth recalling that periods do not appear as input data
in the application of the LM method: the determination of absolute 
magnitudes is totally period-independent.
These results are thus very satisfactory, since oxygen-rich Miras in 
the Large Magellanic Cloud are known to obey a PL relation in the K 
band (Feast et al.\ 1989) with the same slope as ours. 
As far as we know, this is the first time that the existence of a PL 
relation in 104 is demonstrated. This result confirms that 104 is as 
suitable as K to study red variables.\\
The PL fits presented in Eq.~(1-4) are the relations {\it of our 
samples}. The PL relations of the whole oxygen-rich Miras population
are expected to be less luminous due to the effects of the Malmquist
bias (Malmquist 1936).

\subsection{Effects of metallicity on the period--luminosity relation}

It has been discussed for a long time as to whether metallicity effects
in Miras might generate different PL relations.
Wood (1990), extrapolating results from pulsating theory, argues that 
local Miras are intrinsically fainter than in the LMC, because of the 
different metal abundance: Galactic Miras might be about 0.25 mag 
fainter in K than those in the LMC. 
Whitelock et al.\ (1994) showed on the contrary that Miras in the LMC, 
the Galactic globular clusters and the solar neighbourhood might obey
a single PL relation.\\
Recently, van Leeuwen et al.\ (1997) have calibrated the zero-point
of the $M_{\iK}$--$P$ relation for Galactic oxygen-rich Miras by using 
HIPPARCOS parallaxes and adopting the slope of the LMC relation.
They obtained: 
\begin{equation}
  M_{\iK} = 0.94 - 3.47 \log P
\end{equation}
This relation is shown in Fig.~3. Its slope is in very good agreement
with ours. This is a very remarkable result: we find that the slopes
of the Galactic PL relations in K are the same as the LMC one.\\

From the agreement between the distance modulus of the LMC derived 
from their zero-point calibration of the PL relation, and the current 
Cepheid distance modulus (Feast 1995), Van Leeuwen et al.\ conclude 
that it is very unlikely that Miras could be affected by some 
metallicity effects.
Figure~5 is basically the same as Fig.~3, with the addition of the
location of the 11 Miras used by van Leeuwen et al.\ for calibrating 
the zero-point. They are represented by filled circles; the 1$\sigma$ 
error bars are shown.
Among these 11 Miras, 6 belong to our group~1, and 3 to group~3 
(the two others do not appear in our K sample).
Diamonds indicate their absolute K magnitude as determined by the 
LM method.
The other stars of the sample are represented by dots. The lines of
Fig.~3 are reported.
We may conclude from Fig.~5 that: 
\begin{enumerate}
  \item the scatter of the Miras used by van Leeuwen et al.\ around 
   their PL relation is comparable to ours
  \item since these authors could not separate distinct populations, 
   they found a single PL relation which naturally lies between our two 
   fits, and closer to the relation of the predominant population which 
   is group~1.
\end{enumerate}
By this, we stress the necessity to consider large sample of 
Galactic oxygen-rich Mira variables and to separate the different 
populations with the help of a metallicity estimator or with a method 
such as LM when deriving PL relations for the Galaxy.

\begin{figure}
  \centerline{\psfig{figure=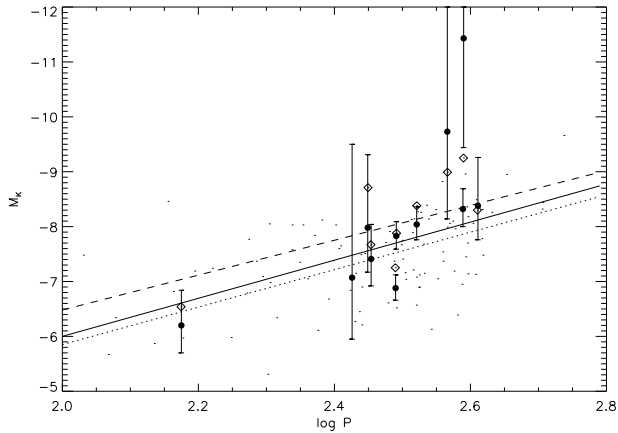,width=3.4in}}
  \caption{Period--luminosity relations in K band. The lines have 
   the same meaning as in Fig.~3. Filled circles are the Miras 
   used by van Leeuwen et al.\ (1997) to calibrate the zero-point of
   the Galactic PL relation and diamonds are their $M_{\iK}$ value as 
   determined by the LM method when available. The other sample stars 
   are represented by dots}
\end{figure}

The results of the present work tend to demonstrate that Galactic 
Miras follow different PL relations, both in K and in 104. The slopes
are the same. Only the zero points differ by about 0.5 mag in K and 
0.8 mag in 104 according to the two distinct populations that we 
have separated. This shift can be interpreted as a metallicity and
mass effect.\\
As mentionned above, according to Wood (1990), changing $Z$ from
$Z_{\odot}$ to $Z_{\odot}$/4 will shift $M_{\iK}$ by 0.25 mag.
Assuming that the magnitude is related to the metallicity by a power
law, a shift by 0.5 mag in K corresponds to a metallicity
of $Z_{\odot}$/16 for the Miras of the E/T disk--halo population,
which value is not striking.

\section{Conclusions}

In this work we have made use of a powerful tool -- the 
maximum-likelihood LM method -- applied to HIPPARCOS data and 
near-infrared apparent magnitudes for a large sample of Mira 
variables.
In K as well as in 104, we separate three populations which exhibit
different kinematics and exponential scale heights.
The two most significant populations can be interpreted as the old
disk population and the extended/thick disk--halo population 
respectively. So they probably differ by their metallicities.
They also exhibit clearly distinct period distribution: this
important result corroborates that the two populations are
certainly composed of stars of different masses, age and metallicity
abundance.\\
The LM method enables us to derive individual distances. They were 
compared to the ones derived from visual magnitudes 
(Mennessier et al.\ 1997a,c).
Distances $r_{\iK}$ are in good agreement with those obtained with
visual data. There is a discrepancy found between the distances 
$r_{104}$ and the distances $r_{\iV}$, which might be due to
sampling effects.\\
Two parallel period--luminosity fit lines are obtained in K as 
well as in 104 for the Mira variables samples. The slope in K is 
very similar to the one observed in the LMC (Feast et al.\ 1989). 
The  Galactic PL relation calibrated by van Leeuwen et al.\ 
(1997) lies between our two fits. The shift between our PL relations 
is probably due to metallicity and mass effects.
We stress the necessity to take into account a possible distinction 
between populations when deriving PL relations for the Galaxy.

\begin{acknowledgements}
We thank F.\ Figueras and J.\ Torra for fruiful and interesting
discussions.
This work was supported by the PICASSO program PICS 348 and by 
the CICYT under contract ESP95-0180.
J.A.M.\ gratefully acknowledges the NASA Grant NAGW-1493 and 
CNRS/GdR 051 grant which partially supported her collaboration.
\end{acknowledgements}

\end{document}